# Heat Evacuation from Active Raman Media Using Quasi-PT Symmetry Coupling

GALINA NEMOVA[1,*] AND CHRISTOPHE CALOZ[1,2]

[1]*Polytechnique Montréal, 2500 ch. de Polytechnique, Montréal, H3T 1J4, QC, Canada.*
[2] *KU Leuven, Kasteelpark Arenberg 10, box 2444, 3001 Leuven, Belgium.*
*Corresponding author: galina.nemova@yahoo.ca

We propose to use frequency-selective quasi parity-time symmetry to reduce the heat generated by coherent anti-Stokes Raman scattering (CARS) reversed cycles in active Raman media with phase mismatch. This is accomplished using a coupled-waveguide structure, which includes the active Raman waveguide (RW), where the Stokes signal undergoes amplification via stimulated Stokes Raman scattering (SSRS), and a dissipative waveguide (DW), which is tuned to the anti-Stokes wavelength so as to evacuate the corresponding anti-Stokes photons from the RW by coupling. The DW introduces optical loss that partially offsets the growth of the anti-Stokes signal in the RW and hence suppress the reversed CARS cycles that would otherwise result into heat generation in the RW. It is shown that the frequency-selective quasi parity-time symmetry provided by the DW can reduce the heat in active Raman media by a very factor of up to five when the CARS phase mismatch is compensated for by the optimum level of coupling between the RW and the DW.

In 1929, Pringsheim proposed to use anti-Stokes fluorescence to cool sodium vapor by leveraging the quantum defect existing between the pump and anti-Stokes photons to annihilate phonons [1]. This approach was later extended to cool solids, such as rare-earth (RE) doped glasses [2] or crystals [3, 4] and to mitigate heat generation in RE-doped lasers or Raman lasers [5-9]. In these cases, each cooling cycle includes the absorption of the pump photon followed by phonon annihilation and radiation of an anti-Stokes photon that removes energy from the system. The energy difference between the anti-Stokes and pump photons is equal to the suppressed heat energy. Unfortunately, a reverse cycle that results into heat generation also takes place in the system. The reversed cycle consists of in anti-Stokes photon absorption followed by pump photon emission and phonon generation that reintroduces some energy into the system. These reverse cycles are major obstacles to laser cooling with anti-Stokes fluorescence, and solutions to this issue are needed to advance this technology.

Parity-time (PT) symmetry, given its powerful capabilities in finely balancing loss and gain, seems to be a natural potential candidate to address this issue. The concept of PT symmetry, or space-time reflection symmetry, was introduced by Bender and Boettcher in 1998 as a complex generalization of quantum theory [10]. It has then opened new perspectives in optics, where it has triggered the development of non-Hermitian photonics [11] and led to new devices, including PT lasers [12-15], optomechanical oscillators [16], unidirectional cloaks [17], microlasers emitting in an angular momentum mode [18], supersymmetric laser arrays [19], and so on.

In this work, we propose to use frequency-selective "quasi-PT" symmetry to suppress the reversed cycles in laser cooled systems, and hence enhance the cooling efficiency. By "quasi-PT", we mean that the symmetry between loss and gain is not complete but offset by a complex coupling mechanism. This approach may represent a major advance in the long-lasting issue of laser cooling limitation due to reverse anti-Stokes cycles in different solids. We shall consider its application to Raman active media and specifically show how frequency-selective quasi-PT symmetry can improve heat mitigation in Raman amplifiers.

The operation of Raman amplifiers is based on stimulated Stokes Raman scattering (SSRS), which is depicted in Fig. 1(a). SSTS is a third-order nonlinear, inelastic scattering process occurring in certain materials, called Raman media, such as tungstate (BaWO$_4$ [20], SrWO$_4$ [21, 22]), nitrate crystals (Ba(NO$_3$)$_2$) [23], vanadate (GdVO$_4$) [24], silicon [25], diamond [26], etc. As all conventional lasers and amplifiers, Raman amplifiers suffer from heat generation within the active medium caused by the quantum defect between the pump and lasing (Stokes) photons, which deteriorates the device performance. In addition to SSRS, such media involve the processes of coherent anti-Stokes Raman scattering (CARS) and stimulated anti-Stokes Raman scattering (SARS), which are respectively represented in Figs. 1(b) and 1(c). SSRS and SARS are processes that generate a single phopon and a single phonon per cycle. In contrast, CARS is a four-wave mixing process that exchanges energy with the medium; causing the annihilaton of two phonon per cycle, and may be used for heat reduction [27].

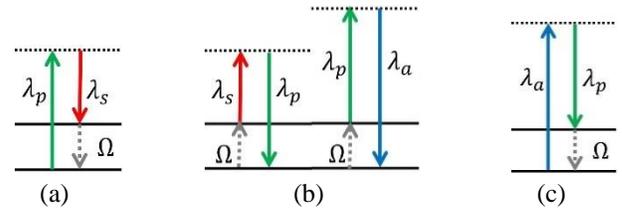

Fig. 1. Processes taking place in a Raman medium: (a) SSRS, (b) CARS, (c) SARS. $\lambda_p$, $\lambda_s$, $\lambda_a$ are the pump, Stokes and anti-Stokes wavelengths, while $\Omega$ is the phonon frequency.

Our enhanced-cooling technique combining anti-Stokes and quasi-PT symmetry is depicted in Fig. 2. Specifically, this technique is based on a *frequency-selective* quasi-PT symmetry and mitigates heat in Raman active media even in the typical presence of some phase mismatch. Contrary to conventional planar Raman lasers and amplifiers, which involve a single waveguide, our structure consists of two coupled waveguides. One of these waveguides is the Raman waveguide (RW), which supports nonlinear pump ($\lambda_p$) power depletion and generation and amplification of the Stokes ($\lambda_s$) and anti-Stokes ($\lambda_a$) modes, while the second waveguide is a dissipative waveguide (DW) that is tuned to the anti-Stokes wavelength to evacuate the related heat energy. The RW-DW coupled pair is designed to form a quasi-PT symmetric system where the DW provides an optimal evacuation channel for the anti-Stokes phonon and an efficient strategy to reduce heat. PT symmetry generally implies an exact balance between gain and loss in photonics, using for instance two identical coupled waveguides with one providing an amount of loss that exactly compensates for the gain of the other. But our amplifier system is more complex than previously reported PT-symmetric photonic systems, and will be shown to require not exact PT symmetry, but only a "quasi-PT" symmetry offset. In our scheme "quasi-PT" symmetry, which is provided by a loss channel made of the DW for the growing anti-Stokes signal in the RW, prevents revers CATS cycles in the RW accompanied by phonon generation.

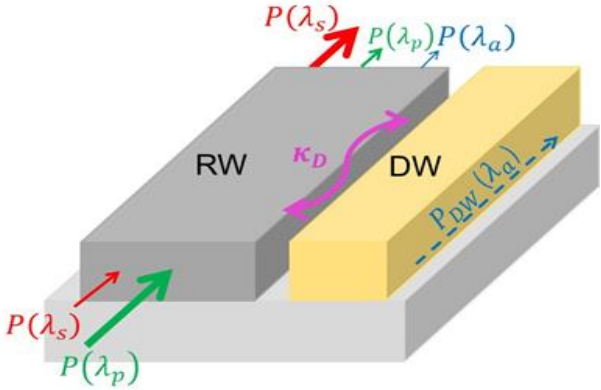

Fig. 2 Proposed quasi-PT symmetric structure to enhance the cooling efficiency of a Raman amplifier, composed of a Raman waveguide (RW), where the generation of the Stokes and anti-Stokes signals takes place, and a dissipative waveguide (DW), that is tuned to the anti-Stokes wavelength to evacuate and dissipate the anti-Stoke photons. $\kappa_D$ is the coupling coefficient between the waveguides.

The operation of the proposed structure can be analyzed with the propagation equations, which describe the evolution of the number of photons in the pump $(N_p)$, Stokes $(N_s)$, and anti-Stokes $(N_a)$ modes propagating in the RW as well as the evolution of the photon number at the anti-Stokes wavelength in the mode propagating in the DW $(N_D)$ along the axis (z) of the coupled structure:

$$\frac{\partial N_p}{\partial z} = \frac{A_{eff}^{RW}}{\mu_0 hc^2} \lambda_p \left[ -\frac{\lambda_s}{\lambda_p} G_s |E_s|^2 |E_p|^2 + G_p |E_a|^2 |E_p|^2 - \alpha_p |E_p|^2 \right], \tag{1}$$

$$\frac{\partial N_s}{\partial z} = \frac{A_{eff}^{RW}}{\mu_0 hc^2} \lambda_s \left[ G_s |E_s|^2 |E_p|^2 + \frac{\lambda_a}{\lambda_s} \widetilde{C_{sa}} - \alpha_s |E_s|^2 \right], \tag{2}$$

$$\frac{\partial N_a}{\partial z} = \frac{A_{eff}^{RW}}{\mu_0 hc^2} \lambda_a \left[ -G_p \frac{\lambda_p}{\lambda_a} |A_a|^2 |E_p|^2 - \widetilde{C_{sa}} + \widetilde{C_{RD}} - \alpha_a |A_a|^2 \right], \tag{3}$$

$$\frac{\partial N_D}{\partial z} = \frac{A_{eff}^{DW}}{\mu_0 hc^2} \lambda_a \left[ -\widetilde{C_{RD}} - \alpha_D |A_D|^2 \right], \tag{4}$$

where $A_{a,D} = E_{a,D} e^{-i\kappa_D}$, $E_p, E_s, E_a$ and $E_D$ are the complex amplitudes of the pump, Stokes, anti-Stokes and DW modes, respectively, $\kappa_D$ is the coupling coefficient between the RW and the DW, $\alpha_p, \alpha_s$, and $\alpha_a$ are the absorption coefficients in the RW at the pump, Stokes, and anti-Stokes wavelengths, respectively, $\alpha_D$ is the optical loss in the DW at the anti-Stokes wavelength, $A_{eff}^{RW}$ and $A_{eff}^{DW}$ are the effective areas of the RW and DW, respectively, and

$$G_s = \frac{1}{4}\sqrt{\frac{\varepsilon_0}{\mu_0}} g_R, \quad G_p = \frac{\lambda_s}{\lambda_p} G_s, \quad C_{sa} = \frac{\lambda_s}{\lambda_a} G_s, \text{ and}$$

$$\widetilde{C_{sa}} = Re(C_{sa} E_p^2 A_a^* E_s^* e^{i(\Delta k - \kappa_D)z}), \widetilde{C_{RD}} = Re(i\kappa_D A_D A_a^*), \tag{5}$$

where $g_R$ is the Raman gain coefficient.

Let us consider the different terms in equations (1)-(4) in some details. The terms with $G_s$ and $G_p$ are related to the process of stimulated Raman scattering. The first terms in the square brackets of equations (2) and (3) describe the change in the number of the Stokes and anti-Stokes photons propagating in the RW as a result of the SSRS and SARS processes, respectively. Both these processes are accompanied by phonon generation (Figs. 1a and 1c). Each of the SSRS or SARS cycles generates a single phonon with the energy $h\Omega$ resulting into heating of the RW. The terms $\widetilde{C_{RD}}$ in equations (3) and (4) describe the change in the anti-Stokes photon numbers in the RW and DW caused by the mutual coupling of these waveguides. The terms $\widetilde{C_{sa}}$ in (2) and (3) are related to the CARS process (Fig. 1b). They include the phase mismatch of the CARS process $\Delta \vec{k} \cdot \vec{z} = (2\vec{k_p} - \vec{k_s} - \vec{k_a}) \cdot \vec{z}$, which depends on the wavevectors of the pump $(\vec{k_p})$, Stokes $(\vec{k_s})$, and anti-Stokes $(\vec{k_a})$ modes. The CARS process is phase matched if $\Delta k = |\Delta \vec{k}| = 0$. The terms $\widetilde{C_{sa}}$ include the phase change $\kappa_D z$ that is caused by the coupling between the RW and the DW. This phase change can compensate for the phase mismatch $\Delta \vec{k} \cdot \vec{z}$. As one can infer from Fig. 1b, the CARS process can take place only in an excited Raman medium because it is triggered by photons that are generated by SSRS. As can be seen in equation (3), the CARS process alternatively serves as a source for anti-Stokes photon generation in the RW.

If the RW and the DW are not coupled ($\kappa_D = 0$) and phase matching is insufficient to satisfy the relation $|\Delta k| \ll G_s |E_p|^2$ (see the Supplement Material), the coefficient $\widetilde{C_{sa}}$ oscillates and changes its sign along the length of the sample. As we already mentioned equation (3) describes the increase $(\widetilde{C_{sa}} < 0)$ or decrease $(\widetilde{C_{sa}} > 0)$ in the anti-Stokes phonon number along the length of the sample as a result of CARS. The decrease in the anti-Stokes phonon number is caused by reversed CARS cycles when all the arrows in Fig. 1b reverse their direction. This reverse CARS process, which starts with the anti-Stokes photon absorption, generates thus a Stokes

photon and two phonons. These phonons are sources of heat and therefore alter the cooling efficiency of the system. The role of the DW is to remove these CARS-generated anti-Stokes photons by the introduction of loss. But the problem is not trivial. If the DW is lossless ($\alpha_D = 0$), the anti-Stokes signal oscillates between the RW and the DW along the coupled structure with a period corresponding to the coherence length. The loss ($\alpha_D \neq 0$) in the DW prevents such oscillation, but it must still be small enough to provide a sufficient coupling length for the initial transfer of the anti-Stokes phonons into it. The loss of the DW should therefore be tune to the optimum accounting for these two antagonistic effects, the ultimate goal being to maximize the amount phonon power dissipation in the RW. This power can be estimated as the difference between the phonon power generated by SSRS plus SARS and the phonon power annihilated or generated by CARS as

$$P_{phon} = h\Omega \frac{A_{eff}^{RW}}{\mu_0 h c^2} \lambda_s G_s \left[ |E_p|^2 (|E_s|^2 + |A_a|^2) - 2Re(E_p^2 A_a^* E_s^* e^{i(\Delta k - \kappa_D) z}) \right], \quad (6)$$

where $h\Omega = hc(\lambda_a^{-1} - \lambda_p^{-1})$ is the phonon energy.

As an example, let us consider a diamond sample operating as a Raman amplifier. The RW is pumped at the wavelength $\lambda_p = 1064$ nm. The Raman gain coefficient in diamond at 1064 nm is $g_R \approx 10$ cm/GW. The wavelengths of the Stokes and anti-Stokes signals are $\lambda_s = 1240$ nm and $\lambda_a = 888$ nm, respectively. The absorption loss in the sample at the Stokes and anti-Stokes wavelengths are $\alpha_s = \alpha_a \approx 0.11$ cm$^{-1}$ [28]. The length of the sample considered in our simulations is $L = 25$ cm. The power of the pump ($P_p$), Stokes ($P_s$), and anti-Stokes ($P_a$) modes propagating in the RW and the power of the waveguide mode propagating in the DW at the anti-Stokes wavelength ($P_D$) are related to the photon numbers in these modes as $P_{p,s,a,D} = hcN_{p,s,a,D}/\lambda_{p,s,a,D}$, where $h$ is the Planck constant and $c$ is the speed of light in vacuum. $N_{p,s,a,D}$ is the photon number in the pump ($p$), Stokes ($s$), anti-Stokes ($a$), and in the waveguide mode of the DW ($D$), respectively. The forthcoming numerical results simulate the different power distributions using equations (1) – (4) while equation (6) to simulate the phonon distribution along the structure.

Let us first consider the case where the RW and the DW would not be coupled ($\kappa_D = 0$) and the CARS process in the RW would be phase matched ($\Delta k = 0$). For simplicity, we define the normalized phase parameter $\widetilde{\Delta} = \Delta k L$. The solid curves in Fig. 3 show the power distributions of the pump, Stokes and anti-Stokes signals in solid lines along the RW. The pump power decreases along the sample, due to power transfer to the Stokes and anti-Stokes signals, and the powers of the Stokes and anti-Stokes signals are almost equal to each other. The solid curve in Fig. 4 shows the corresponding phonon power variation. We see that in the phase matched case, where the reverse CARS cycles are absent, the CARS process annihilates most of the phonons generated by SSRS and SARS.

However, such a matching condition is typically unrealistic in practice. Let us therefore introduce some phase mismatch into the system. For example, let us consider a system with the normalized phase mismatch $\widetilde{\Delta} = 25$. The evolution of the powers in the pump, Stokes, and anti-Stokes signals along this mismatched RW are plotted in dotted lines in Fig. 3. Now, the powers of the Stokes and anti-Stokes signals are very different from each other for most structure lengths. The reverse CARS cycles together with the dramatical growth of the Stokes signal result to the dramatical growth of the phonon power in the RW, as shown by the dotted curve in Fig. 4.

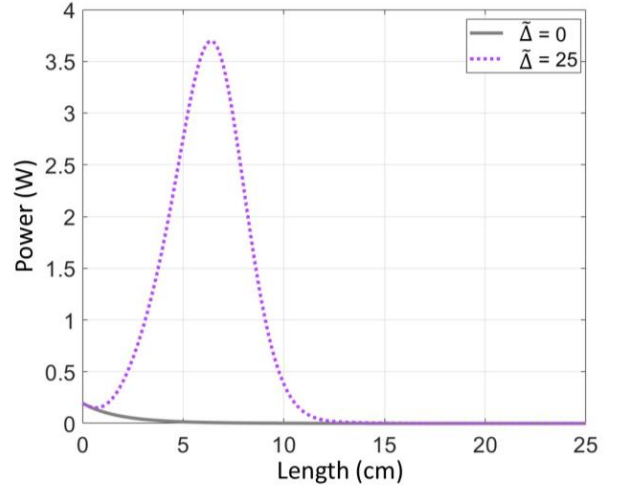

Fig. 4. Phonon power distribution along the RW without coupling ($\kappa_D = 0$) and with phase matching ($\widetilde{\Delta} = 0$) and phase mismatch ($\widetilde{\Delta} = 25$).

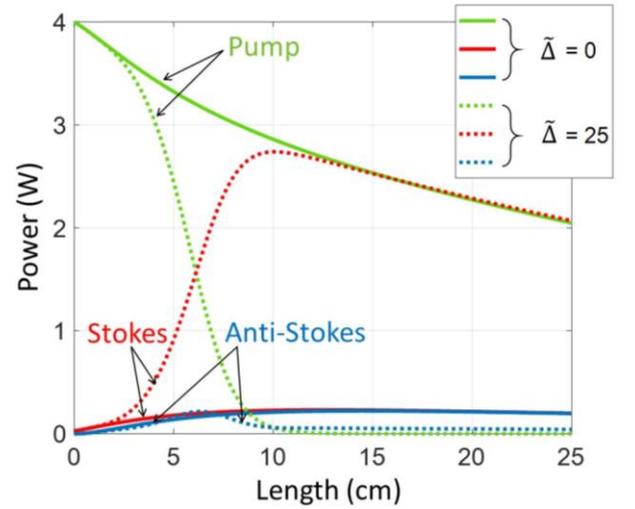

Fig. 3. Power distribution for the pump, Stokes and anti-Stokes signals along the length of the RW without coupling ($\kappa_D = 0$), for the case of phase matched ($\widetilde{\Delta} = 0$) and phase mismatched ($\widetilde{\Delta} = 25$).

As one can see in equation (5), the phase mismatch in the *system* can be compensated for by introducing coupling, $\kappa_D \neq 0$. For simplicity, and consistency with the normalized phase, we define the normalized coupling constant $\widetilde{\kappa_D} = \kappa_D L$. Let us consider a system with $\widetilde{\kappa_D} = 25$, so as to fully compensate for the phase mismatch $of\ \widetilde{\Delta} = 25$. As one can see in Fig. 5, major heat suppression is achieved in this case, the phonon power being decreased from ~3.7 W (dashed curve) to ~0.55 W (solid curve). Unfortunately, realizing coupled waveguides providing such perfect compensation is not a simple task.

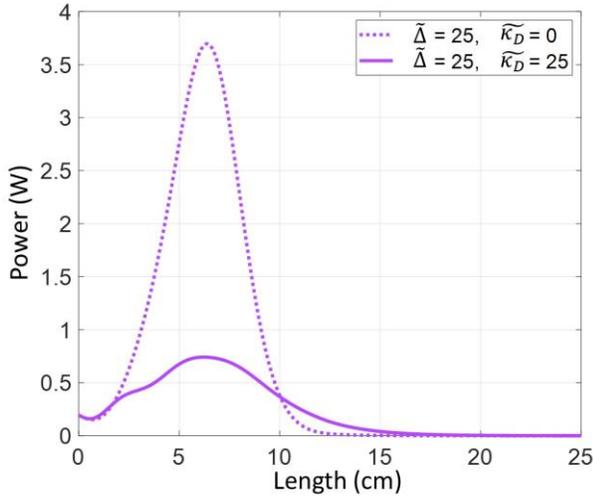

Fig. 5. Phonon power distribution along the RW without coupling ($\widetilde{\kappa_D} = 0$) and with coupling ($\widetilde{\kappa_D} = 25$) and with phase mismatch ($\widetilde{\Delta} = 25$).

Let us therefore consider the more realistic system where the phase mismatch can be only partially compensated for by coupling. As an example, we set $\widetilde{\Delta} = 25$ and $\widetilde{\kappa_D} = 40$, and we shall compare two designs, one with DW free from any optical loss, i.e., $\widetilde{\alpha_D} = \alpha_D L = 0$, and one with an of optical loss $\widetilde{\alpha_D} = 25$. The power distributions along this structure are plotted in Fig. 6. In the absence of DW loss ($\widetilde{\alpha_D} = 0$), the power of the anti-Stokes signal is higher than that with loss. Indeed, in the former case, the waveguide coupling constant $\widetilde{C_{RD}}$ oscillates between positive and negative values along the structure, as shown by the solid curve in Fig. 7. Unfortunately, these recoupled anti-Stokes photons can cause reverse CARS and SASR, resulting into the heat generation in the RW. The phonon power generated in the RW when the optical loss in the DW are absent is shown again, for comparison, in solid line in Fig. 8, with a peak of ~3.7 W (factor 2.4 or decrease of 56.7%).

The situation with loss is different. Since in that case the power of the anti-Stokes signal decreases along the structure (Fig. 6), the waveguide coupling constant $\widetilde{C_{RD}}$ becomes exclusively negative (Fig. 7). The majority of the anti-Stokes photons propagating in the DW are thus dissipated in the DW before having a chance of returning to the RW and generating heat there via reverse CARS and SARS. As a result, the phonon power in the system has decreased from ~3.7 W to ~2.4 W (factor 1.4 or decrease of 35%), as shown by the dotted curve in Fig. 8.

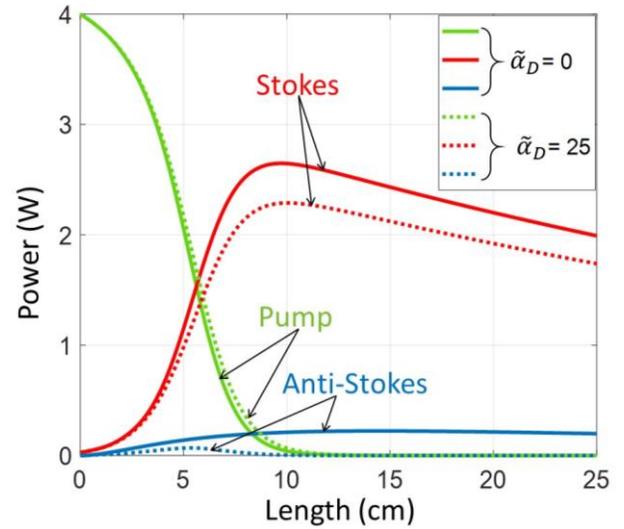

Fig. 6. Power distributions of the pump, Stokes, and anti-Stokes signals along the RW with coupling ($\widetilde{\kappa_D} = 40$) and phase mismatch ($\widetilde{\Delta} = 25$), without loss ($\widetilde{\alpha_D} = 0$) and with loss $\widetilde{\alpha_D} = 25$.

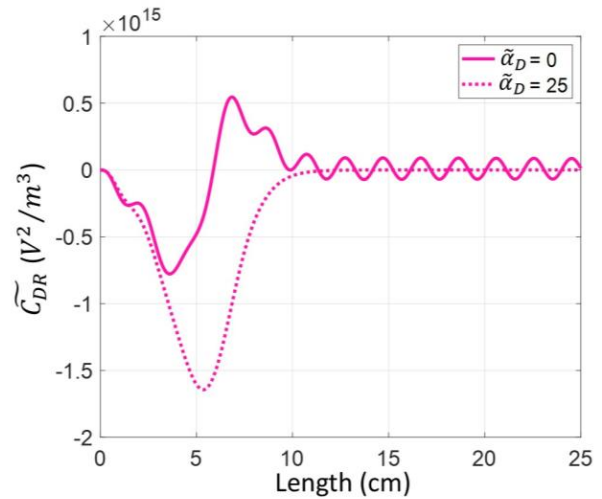

Fig. 7. $\widetilde{C_{DR}}$ parameter along the structure with ($\widetilde{\kappa_D} = 40$) and phase mismatched ($\widetilde{\Delta} = 25$), without loss ($\widetilde{\alpha_D} = 0$) and with loss $\widetilde{\alpha_D} = 25$. Negative and positive values represent respectively phonon annihilation and generation.

Finally, Fig. 9 represents the evolution of the system performance in terms of the DW loss. While the phonon power reduction monotonically increases with increasing DW loss, this loss unavoidably also affects the power level of the Stokes photons, and hence the power level of the overall device. If stability is the primary concern in the application considered, one may introduce a large amount of loss in the DW (e.g., here $\widetilde{\alpha_D} = 30$); if, in contrast, in the output power level is also an important parameter, one may tradeoff some stability for higher power by operating in a lower loss regime (e.g. here $\widetilde{\alpha_D} = 5$).

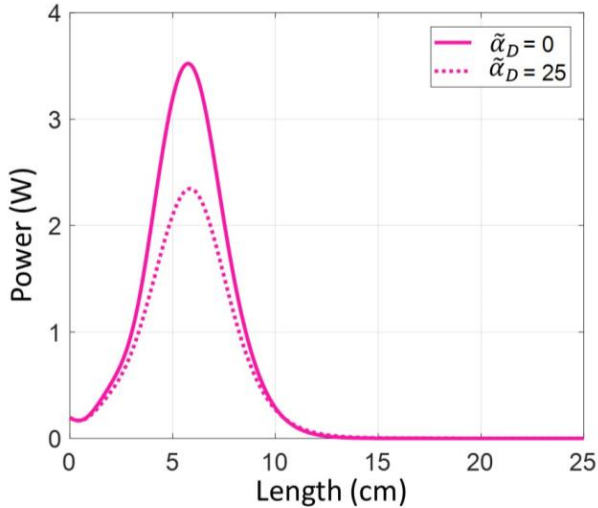

Fig. 8. Phonon power distribution along the RW with coupling $\widetilde{\kappa_D} = 40$ and phase mismatch $\widetilde{\Delta} = 25$, without loss ($\widetilde{\alpha_D} = 0$) and with loss $\widetilde{\alpha_D} = 25$.

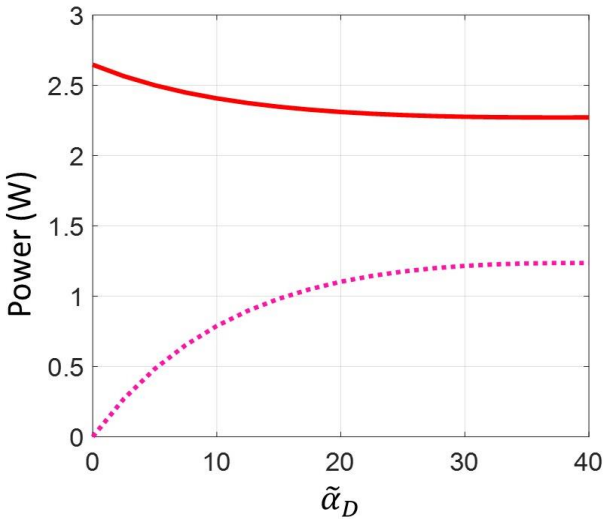

Fig. 9. Evolution of the system performance versus the DW loss ($\widetilde{\alpha_D}$) for coupling $\widetilde{\kappa_D} = 40$ and phase mismatch $\widetilde{\Delta} = 25$. The dotted curve represents the difference in the phonon peak power in the RW without loss and with loss (Fig. 8 at $\widetilde{\Delta} = 25$), while the solid curve represents the Stokes power in the RW.

In summary, we have introduced the concept of frequency-selective quasi-PT symmetry for cooling active Raman waveguide systems by coupling a dissipative waveguide to the amplification waveguide for optimal phonon suppression. We have illustrated this concept with the example of Stokes amplification in Raman amplifiers, where heat reduction by a factor of more than two has been demonstrated in the case of a phase matched system amplifier. This concept represents a novel approach of anti-Stokes cooling and may play a significant role in the development of more stable and athermal amplifiers and lasers.

## References

1. P. Pringsheim, Z. Phys. 57, 739 (1929).
2. J. V. Guiheen, C. D. Haines, G. H. Sigel, R. I. Epstein, J. Thiede, W. M. Patterson, Phys. Chem. Glasses Eur. J. Glass Sci. Technol. B **47**, 167 (2006)
3. S. Rostami, A. R Albrecht, A. Volpi, M. Sheik-Bahae, Photonics Research, 7, 445 (2019).
4. E. Soares de Lima Filho, K. V. Krishnaiah, Y. Ledemi, Ye-Jin Yu, Y. Messaddeq, G. Nemova, R. Kashyap, Opt. Express, 4630 (2015).
5. G. Nemova, R. Kashyap, J. Lightwave Technol., 27, 5597 (2009).
6. G. Nemova, R. Kashyap, J. Opt. Soc. Am. B, 26, 2237 (2009).
7. E. Mobini, M. Peysokhan, B. Abaie, A. Mafi, J. Opt. Soc. Am. B, 35 (10), 2484 (2018).
8. M. Peysokhan, E. Mobini, A. Allahverdi, B. Abaie, A. Mafi, Photonics Research, 8(2), 202 (2020).
9. N. Vermeulen, C. Debaes, P. Muys, H. Thienpont, Phy. Rev. Lett., **99**, 093903 (2007).
10. C. M. Bender and S. Boettcher Phys. Rev. Lett. 80, 5243 (1998).
11. K. G. Makris, R. El-Ganainy, D. N. Christodoulides and Z. H. Musslimani, Phys. Rev. Lett. **100**, 103904 (2008).
12. M. Brandstetter, M. Liertzer, C. Deutsch, P. Klang, J. Schoberl, H. E. Tureci, G. Strasser, S. Unterrainer Kand Rotter, Nat. Commun. **5**, 4034 (2014).
13. B. Peng, S. K. Özdemir, S. Rotter, H. Yilmaz, M. Liertzer, F. Monifi, C. M. Bender, F. Nori and L. Yang, Science, **346**, 328 (2014).
14. H. Hodaei, M. A. Miri, M. Heinrich, D. N. Christodoulides and M. Khajavikhan Science, **346**, 975 (2014).
15. L. Feng, Z. J. Wong, R. M. Ma, Y. Wang and X. Zhang, Science, **346**, 972 (2014)
16. J. Wiersig, Phys. Rev. Lett., **112**, 203901 (2014).
17. Z. Lin, H. Ramezani, T. Eichelkraut, T. Kottos, H. Cao and D. N. Christodoulides, Phys. Rev. Lett. **106**, 213901 (2011).
18. P. Miao, Z. Zhang, J. Sun, W. Walasik, S. Longhi, N. M Litchinitser, L Feng, Science, **353**, 464 (2016).
19. M. Teimourpour, L. Ge, D. N. Christodoulides, R. El-Ganainy, Sci. Rep. **6**, 33253 (2016).
20. X. H. Chen, P. Li, X. Y. Zhang, Q.P. Wang, Z.J. Liu, Z.H. Cong, L. Li, H.J Zhang, Laser Phys., **21**, 2040 (2011).
21. H. Zhang, P. Li, Q. Wang, Q. Wang, X. Chen, X. Zhang, J. Chang, and X. Tao, Appl Optics, **53**, 7189 (2014).
22. H. Zhang H, P. Li, Appl. Phys. B, **122**,12 (2016).
23. P. V. Shpak, S. V. Voitikov, R. V. Chulkov, P. A. Apanasevich, V. A. Orlovich, A. S. Grabtchikov, A. Kushwaha, N. Satti, L. Agrawal, A. K. Maini, Opt. Commun., **285**, 3659 (2012).
24. M. Frank, S. N. Smetanin, M. Jelínek Jr., D. Vyhlídal, V. E. Shukshin, P. G. Zverev, V. Kubeček, Laser Phys. Lett., **16**, 085401 (2019).
25. T. Yasuda, M. Okano, M. Ohtsuka, M. Seki, N. Yokoyama, Y. Takahashi, OSA Continuum, **3**, 814 (2020).
26. K. Chrysalidis, V. N. Fedosseev, B. A. Marsh, R. P. Mildren, D. J. Spence, K. D. A. Wendt, S. G. Wilkins, E. Granados, Opt. Lette., 44, 3924 (2019).
27. B. Bobbs, C. Warner, J. Opt. Soc. Am. B, **7**, 234 (1990).
28. Y. Li, Z. Bai, H. Chen, D. Jin, X. Yang, Y. Qi, J. Ding, Y. Wang, Z. Lu, Results in Physics, **16**, 192853 (2020).

# Supplement Material

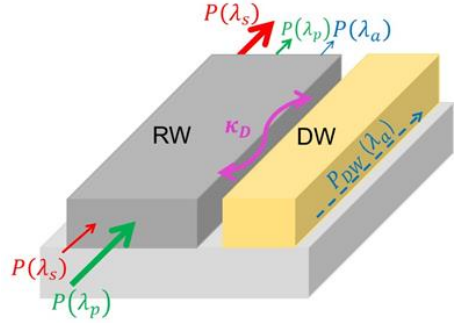

Fig. 1 The structure under investigation. It includes a RM, where generation of the Stokes and anti-Stokes signals take place, and a DW, which support propagation and introduces losses only at the anti-Stokes wavelength. $\kappa_D$ is the coupling coefficient.

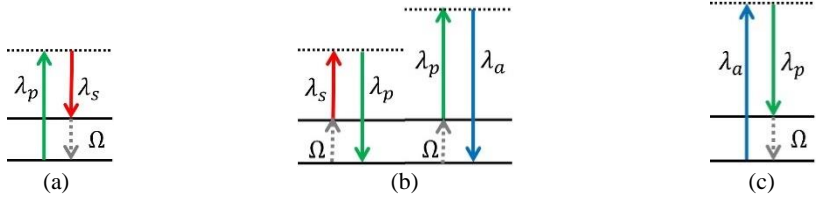

Fig. 2. Process, which takes place in a Raman medium: (a) SSRS, (b) CARS, (c) SARS. $\lambda_p$, $\lambda_s$, $\lambda_a$ are the pump, Stokes, anti-Stokes wavelengths, $\Omega$ is the phonon frequency.

Let us consider the operation an active Raman waveguide (RW), which is pumped at the wavelength $\lambda_p$ and coupled with the dissipative waveguide (DW) as illustrated in Fig. 1. The pumped RW can generate and support the propagation of Stokes (at the wavelength $\lambda_s$) and anti-Stokes (at the wavelength $\lambda_a$) modes. The generation of the Stokes and anti-Stokes photons is based on the processes of stimulated Stokes Raman scattering (SSRS) and coherent anti-Stokes Raman scattering (CARS), respectively (Fig. 2 a & b). The DW can only support a waveguide mode at the anti-Stokes wavelength. It can introduce optical losses $\alpha_D$ at the anti-Stokes wavelength in the system. The operation process in the system can be analyzed with the system of equations, which describes the evolution of the electric field of the pump ($E_p$), Stokes ($E_s$), anti-Stokes ($E_a$), and the DW ($E_D$) modes, respectively.

$$\frac{\partial E_p}{\partial z} = -\frac{\lambda_s}{\lambda_p} G_s |E_s|^2 E_p + G_p |E_a|^2 E_p - \alpha_p E_p, \tag{1}$$

$$\frac{\partial E_s}{\partial z} = G_s |E_p|^2 E_s + \frac{\lambda_a}{\lambda_s} C_{sa} E_p^2 E_a^* e^{i\Delta k z} - \alpha_s E_s, \tag{2}$$

$$\frac{\partial E_a}{\partial z} = -G_p \frac{\lambda_p}{\lambda_a} |E_p|^2 E_a - C_{sa} E_p^2 E_s^* e^{i\Delta k z} + i\kappa_D E_D - \alpha_a E_a, \tag{3}$$

$$\frac{\partial E_D}{\partial z} = i\kappa_D E_D - \alpha_D E_D. \tag{4}$$

Here $G_s = \frac{1}{4}\sqrt{\frac{\varepsilon_0}{\mu_0}} g_R$, $G_p = \frac{\lambda_s}{\lambda_p} G_s$, $C_{sa} = \frac{\lambda_s}{\lambda_a} G_s$, $g_R$ is the Raman gain coefficient, and $\Delta k = |\vec{\Delta k}| = |2\vec{k_p} - \vec{k_s} - \vec{k_a}|$ is the mismatch, where the wavevectors of the pump $(\vec{k_p})$, Stokes $(\vec{k_s})$, and anti-Stokes $(\vec{k_a})$ modes.

If $\Delta k = 0$ the CARS process is phase matched. If $\Delta k \neq 0$ it is mismatched.

Let us consider $E_s(z) = B_s(z)e^{i\Delta kz/2}$, $E_a(z) = B_a(z)e^{i\Delta kz/2}$, and $E_D(z) = B_D(z)e^{i\Delta kz/2}$. In this case equations (1)-(4) will look like

$$\frac{\partial E_p}{\partial z} = -\frac{\lambda_s}{\lambda_p}G_s|B_s|^2 E_p + G_p|B_a|^2 E_p - \alpha_p E_p,$$
(1a)

$$\frac{\partial B_s}{\partial z} = -\left(i\frac{\Delta k}{2} - G_s|E_p|^2 + \alpha_s\right)B_s + \frac{\lambda_a}{\lambda_s}C_{sa}E_p^2 B_a^*,$$
(2a)

$$\frac{\partial B_a}{\partial z} = -\left(i\frac{\Delta k}{2} + G_p\frac{\lambda_p}{\lambda_a}|E_p|^2 + \alpha_a\right)B_a - C_{sa}E_p^2 B_s^* + i\kappa_D B_D,$$
(3a)

$$\frac{\partial B_D}{\partial z} = i\kappa_D B_D - \left(i\frac{\Delta k}{2} + \alpha_D\right)B_D.$$
(4a)

As one can see in the first terms in the brackets (2a) and (3a) the CARS process is quasi phase matched if $|\Delta k| \ll G_s|E_p|^2$.

Let us consider the following relation:

$$Re\left(E^*\frac{\partial E}{\partial z}\right) = \frac{1}{2}\frac{\partial}{\partial z}|E|^2$$
(5)

Using this relation and equations (1) – (4) we can obtain the equations for the evolution of power and the number of photons in the pump, Stokes, anti-Stokes and the DW modess. Indeed,

$$P_{p,s,a} = \frac{1}{2}\left(\frac{\varepsilon_0}{\mu_0}\right)^{\frac{1}{2}}|E_{p,s,a}|^2 A_{eff}^{RW}, \quad \text{and} \quad P_D = \frac{1}{2}\left(\frac{\varepsilon_0}{\mu_0}\right)^{\frac{1}{2}}|E_D|^2 A_{eff}^{DW},$$
(6)

where $A_{eff}^{RW}$ and $A_{eff}^{DW}$ are the effective RW and DW area, respectively. $P_{p,s,a,D}$ are the powers of the pump ($p$), Stokes ($s$), anti-Stokes ($a$) modes and the power of the waveguide mode in the DW ($D$), respectively. The number of photons in the pump $(N_p)$, Stokes ($N_s$), anti-Stokes $(N_a)$ and the waveguide mode of the DW $(N_D)$ can be calculated using the relation

$$N_{p,s,a,D} = \frac{P_{p,s,a,D}}{h\nu_{p,s,a,D}},$$
(7)

where $h$ is the Planck constant, $\nu_{p,s,a,D} = \frac{c}{\lambda_{p,s,a,D}}$ and $\lambda_D = \lambda_a$.

The equations, which describe the evolution of the number of photons in the pump $(N_p)$, Stokes ($N_s$), and anti-Stokes ($N_a$) modes of the RW as well as the evolution of the number of photons at the anti-Stokes wavelength in the mode of the DW ($N_D$) look like:

$$\frac{\partial N_p}{\partial z} = \frac{A_{eff}^{RW}}{\mu_0 hc^2}\lambda_p\left[-\frac{\lambda_s}{\lambda_p}G_s|E_s|^2|E_p|^2 + G_p|E_a|^2|E_p|^2 - \alpha_p|E_p|^2\right],$$
(8)

$$\frac{\partial N_s}{\partial z} = \frac{A_{eff}^{RW}}{\mu_0 hc^2}\lambda_s\left[G_s|E_s|^2|E_p|^2 + \frac{\lambda_a}{\lambda_s}\widetilde{C_{sa}} - \alpha_s|E_s|^2\right],$$
(9)

$$\frac{\partial N_a}{\partial z} = \frac{A_{eff}^{RW}}{\mu_0 hc^2}\lambda_a\left[-G_p\frac{\lambda_p}{\lambda_a}|A_a|^2|E_p|^2 - \widetilde{C_{sa}} + \widetilde{C_{RD}} - \alpha_a|A_a|^2\right],$$
(10)

$$\frac{\partial N_D}{\partial z} = \frac{A_{eff}^{DW}}{\mu_0 hc^2}\lambda_a\left[-\widetilde{C_{RD}} - \alpha_D|A_D|^2\right],$$
(11)

where $A_{a,D} = E_{a,D}e^{-i\kappa_D}$. $E_p, E_s, E_a$, and, $E_D$ are the complex amplitudes of the pump, Stokes, anti-Stokes and DW modes, respectively. $\kappa_D$ is the coupling coefficient between the RW and the DW. $\alpha_p, \alpha_s$, and $\alpha_a$ are the absorption coefficients in the RW at the pump, Stokes, and anti-Stokes wavelengths, respectively and $\alpha_D$ is the optical losses in the DW at the anti-Stokes wavelength. $A_{eff}^{RW}$ and $A_{eff}^{DW}$ are the effective area of the RW and DW, respectively, and $\widetilde{C_{sa}} = Re(C_{sa}E_p^2 A_a^* E_s^* e^{i(\Delta k - \kappa_D)z})$, $\widetilde{C_{RD}} = Re(i\kappa_D A_D A_a^*)$.

Each SSRS and SARS cycle *generate* a single phonon (Fig. 2 a, c). The first terms in the square brackets in equation (9) and (10) describe the number of phonons generated with SSRS and SARS, respectively. The number of phonons generated in the RW at the $z$ point with both cycles is

$$N_{Phonon}^{(SSRS\_SARS)} = \frac{A_{eff}^{RW}}{\mu_0 h c^2} \lambda_s G_s |E_s|^2 |E_p|^2 + \frac{A_{eff}^{RW}}{\mu_0 h c^2} \lambda_p G_p |E_p|^2 |A_a|^2 = \frac{A_{eff}^{RW}}{\mu_0 h c^2} \lambda_s G_s |E_p|^2 (|E_s|^2 + |A_a|^2). \tag{12}$$

The term $\widetilde{C_{sa}}$ in equations (9) and (10) describe the number of phonons *annihilated* or *generated* in the RW with CARS or revers CARS cycles, respectively. The number of phonons associated with the CARS cycles on at the $z$ point is

$$N_{Phonon}^{(SSR\_SARS\_-CARS)} = 2\frac{A_{eff}^{RW}}{\mu_0 h c^2} \lambda_a C_{sa} Re\left(E_p^2 A_a^* E_s^* e^{i(\Delta k - \kappa_D)z}\right) = 2\frac{A_{eff}^{RW}}{\mu_0 h c^2} \lambda_s G_s Re\left(E_p^2 A_a^* E_s^* e^{i(\Delta k - \kappa_D)z}\right). \tag{13}$$

Here 2 means that each CARS cycle is accompanied by annihilation or generation of *two* phonons. Using relations (12) and (13) we can estimate the phonon power in the system as

$$P_{phon} = h\nu_{phon} \frac{A_{eff}^{RW}}{\mu_0 h c^2} \lambda_s G_s \left[|E_p|^2(|E_s|^2 + |A_a|^2) - 2Re\left(E_p^2 A_a^* E_s^* e^{i(\Delta k - \kappa_D)z}\right)\right], \tag{14}$$

where $h\nu_{phon} = hc(\lambda_a^{-1} - \lambda_p^{-1})$ is the phonon energy.

We want to emphasize that exact PT symmetry, in the sense of completely balanced gain and loss, is not required in our structure; instead, *quasi*-PT symmetry, which can be realized by including a global loss offset, has been implemented.